\documentclass[amsmath,amssymb,aps,twocolumn,pra]{revtex4-1}%,superscriptaddress
\usepackage{graphicx}
\usepackage{soul}\usepackage[colorlinks=true,citecolor=blue,linkcolor=magenta]{hyperref}
\usepackage[usenames]{color}
\usepackage{relsize}
\usepackage[english]{babel}
\usepackage{enumerate}
\usepackage{pifont}
\usepackage{amsfonts}
\usepackage[latin1]{inputenc}
\usepackage{lineno}

\newcommand{\ket}[1]{\left| #1 \right>} % for Dirac bras
\usepackage[normalem]{ulem} %makes underlining possible: \uline \uwave \sout
\definecolor{mgreen}{RGB}{1,123,0}

\newcommand {\grsim} {\ {\raise-.5ex\hbox{$\buildrel>\over\sim$}}\ }
\newcommand {\lessim} {\ {\raise-.5ex\hbox{$\buildrel<\over\sim$}}\ }

\begin{document}
%TC:ignore
%\linenumbers

\title{Tailoring quantum gases by Floquet engineering}

\author{Christof~Weitenberg,} %$^{1,2}$, Juliette~Simonet$^{1,2}$}
\email{christof.weitenberg@physnet.uni-hamburg.de}
\affiliation{ILP -- Institut für Laserphysik, Universität Hamburg, Luruper Chaussee 149, 22761 Hamburg, Germany}
\affiliation{The Hamburg Centre for Ultrafast Imaging, Luruper Chaussee 149, 22761 Hamburg, Germany}
\author{Juliette~Simonet}
\affiliation{ILP -- Institut für Laserphysik, Universität Hamburg, Luruper Chaussee 149, 22761 Hamburg, Germany}
\affiliation{The Hamburg Centre for Ultrafast Imaging, Luruper Chaussee 149, 22761 Hamburg, Germany}

\date{\today}
\maketitle

%%%%%%%%%% 
%Vorpsanntext
%%%%%%%%%%%

\textbf{Floquet engineering is the concept of tailoring a system by a periodic drive. It has been very successful in opening new classes of Hamiltonians to the study with ultracold atoms in optical lattices, such as artificial gauge fields, topological band structures and density-dependent tunneling. Furthermore, driven systems provide new physics without static counterpart such as anomalous Floquet topological insulators. In this review article, we provide an overview of the exciting developments in the field and discuss the current challenges and perspectives.
}

\subsection*{Introduction} 
The philosophy of quantum simulation with ultracold atoms is to engineer the Hamiltonian of interest by adding the relevant terms step by step, e.g., tailored potentials, additional internal states or controlled interactions. A new dimension is opened by time-dependent control of the system, which allows adding terms such as artificial gauge fields or density-dependent tunneling. Modifying a system by periodic driving is called Floquet engineering and has been proven to be a very powerful tool \cite{Bukov2015, Eckardt2017}. While driving a system at particular frequencies has always been a preferred method for probing it, Floquet engineering radically changes the perspective and uses the same technique to modify it. When periodically shaking an optical lattice, instead of performing spectroscopy of the band structure, one can hybridize the bands to produce bands with new physical properties. Beyond such engineering of static Hamiltonians, Floquet driving can also give rise to new phases without static counterpart. A classical analogy of this Floquet idea is Kapitza's pendulum, which has the upright state as a new stable equilibrium position due to a fast drive of the pivot point.

Floquet engineering as a general concept is used in many areas including solid-state physics and synthetic systems such as photonic waveguides \cite{Ozawa2019, Oka2019, Rudner2020} and it stimulates intense exchange between these areas. In cold atom research, this approach is particularly fruitful due to the high control over these systems, which allows implementing a multitude of driving schemes, as well as due to the easily accessible time scales. E.g., driving a solid-state crystal at the relevant scales requires laser beams of extremely high intensity, which can be provided only by pulsed lasers and requires suitably fast measurement schemes \cite{McIver2020}. In cold atoms, the driving frequency is typically in the Kilohertz regime and the displacement amplitude can span over many lattice sites, such that the main limitation is the heating inherently associated with Floquet systems. In this review article, we want to give a short overview of the techniques known as Floquet engineering with a focus on optical lattices and discuss the new physics that has been explored with quantum gas experiments.

\subsection*{Effective Hamiltonian and renormalized tunneling}
Floquet systems are periodically driven systems described by a Hamiltonian $\hat{H}(t+T)=\hat{H}(t)$ with the driving period $T=2\pi/\Omega$. The primary interest in Floquet systems is the fact that despite being time dependent, they can be described by a time-independent effective Hamiltonian $\hat{H}_{\mathrm{eff}}$, if one probes at multiples of the driving period, i.e. the time-evolution operator is given by $U(t_0+T,t_0 )=e^{i\hat{H}_{\mathrm{eff}} T/\hbar}$ \cite{Bukov2015, Eckardt2017}. The dynamics within one driving period, the so-called micromotion, can often be separated from the long-time dynamics. The effective Hamiltonian can be calculated, e.g., by a high-frequency expansion in powers of $1/\Omega$ \cite{Bukov2015, Eckardt2017}. In lowest order, which is a suitable approximation in the high-frequency limit, it is given by the average over one driving period $\hat{H}_{\mathrm{eff}}=<\hat{H}(t)>_T$.  

As a simple but insightful example, let us consider ultracold atoms in a driven one-dimensional optical lattice, i.e. a periodic potential formed by the interference of two laser beams. In the tight-binding description, the dispersion relation is given by the energy $E(k)=-2 J \cos (kd)$ with the tunneling element $J$, the quasimomentum $k$ and the lattice constant $d$. Lattice shaking can be easily realized by periodically changing the frequency of one laser beam and it results in a oscillating inertial force of the form $F_0 \cos (\Omega t)$, i.e. with zero average force. For an atom prepared in a certain quasimomentum state, lattice shaking leads to a periodic oscillation in quasimomentum space. In the high frequency limit, the effective energy is given by the average over the varying energy explored during one oscillation (Fig. 1a). The effective Hamiltonian is now described by a tight-binding dispersion $E_{\mathrm{eff}}$ with a renormalized tunneling element $J_{\mathrm{eff}}=J \mathcal{J}_0 (K_0)$, where $\mathcal{J}_0$ is the Bessel function of the first kind of order zero and $K_0=\frac{d F_0}{\hbar \Omega}$ is the driving strength. This renormalization of $J$ with the Bessel function was mapped out experimentally by studying the rate of expansion of an atomic cloud, which is directly related to $J$ (Fig. 1a)\cite{Lignier2007}. The phenomenon of complete suppression of tunneling at the zero crossing of the Bessel function is known as dynamical localization. The same effect as lattice shaking can also be produced in a static lattice by applying an oscillating external force, e.g. from an oscillating magnetic field gradient \cite{Jotzu2015}.

For sufficiently large driving amplitudes where $\mathcal{J}_0 (K_0)<0$, one obtains an inverted band structure with new minima at the edges of the Brillouin zone. A Bose-Einstein condensate (BEC) in a shaken lattice will recondense at the new minima, which can be directly observed in the quasimomentum distribution obtained by a time-of-flight expansion (Fig. 1a inset). This sign inversion of the tunneling elements is particularly interesting in a triangular lattice, where it can lead to frustration and intriguing magnetic phases of the classical XY model \cite{Struck2011, Struck2013}. An alternative scheme is to dress the lowest band with the first excited band by near-resonantly driving the lattice with the band energy difference \cite{Parker2013, Fujiwara2019}. This produces, e.g., a double-well dispersion in the lowest band, which can be mapped to a ferromagnetic spin model \cite{Parker2013}. Starting from this basic concept of tunnel renormalization, one can realize increasingly complex Hamiltonians by employing more and more sophisticated driving schemes, as we will see in the following.

\begin{figure*}%[t]
\includegraphics[width=0.8\textwidth]{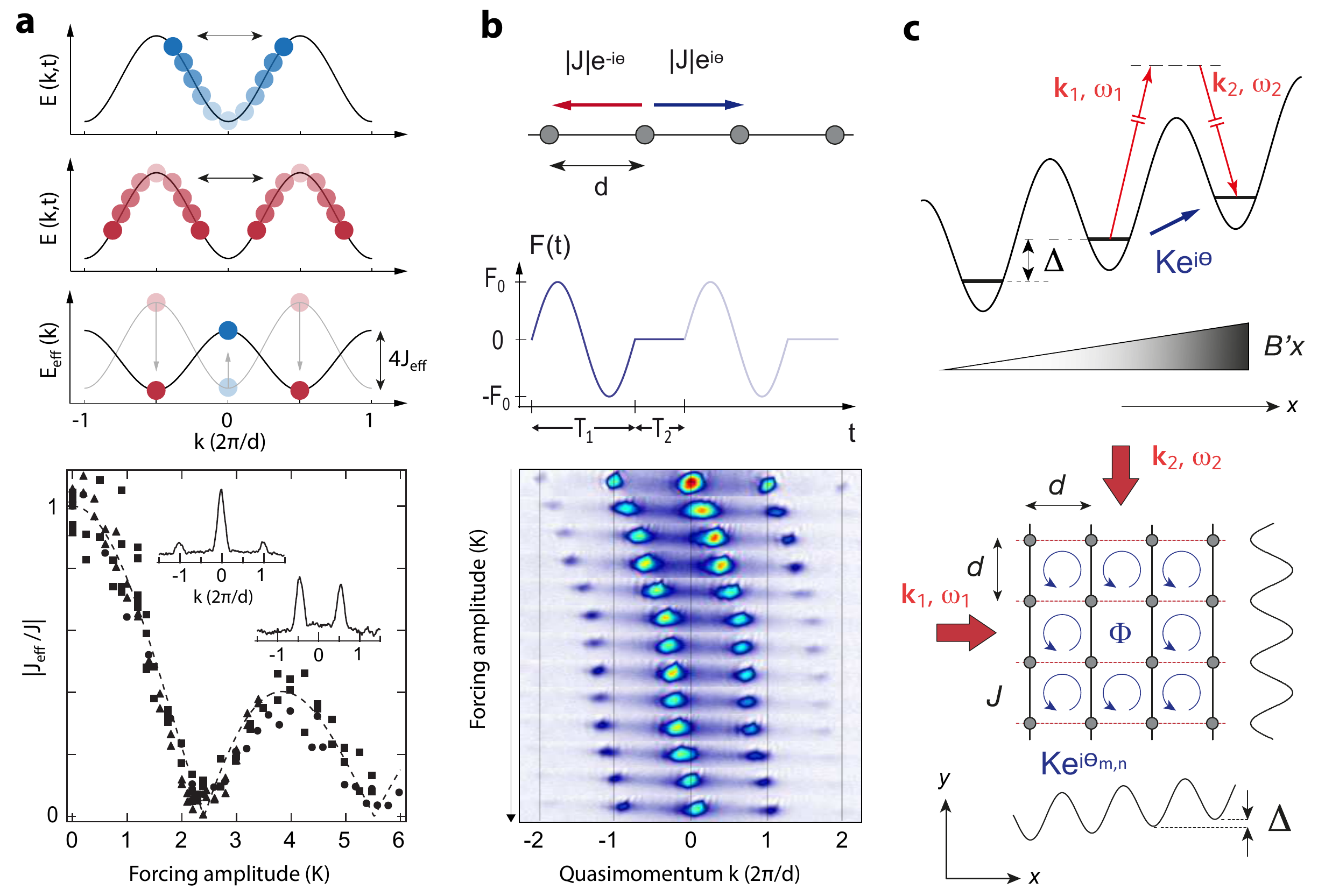}
\vspace{-0.cm}\caption{\textbf{Effective Hamiltonian and renormalized tunneling.} \textbf{a}, In a driven optical lattice, the inertial force leads to an oscillation of the atoms in momentum space and in high-frequency approximation the effective energy is given by the average over one driving period. This leads to an effective dispersion relation described by a renormalized tunneling element $J_{\mathrm{eff}}$, which can become negative for sufficiently strong driving. Bottom: The renormalization by the Bessel function $\mathcal{J}_0$ as measured by the suppressed diffusion of a BEC. Adapted from \cite{Lignier2007}. \textbf{b}, Shaking protocols breaking TRS allow realizing tunable Peierls phases $\theta$ of the tunneling elements beyond 0 and $\pi$. Increasing the driving amplitude $K_0$ continuously shifts the minimum of the lowest Bloch band and gives rise to superfluids at finite quasi-momentum. Adapted from \cite{Struck2012}. \textbf{c}, In an optical lattice tilted, e.g. by a magnetic field gradient $B^{'}$, the tunneling can be reestablished using laser-assisted tunneling imprinting Peierls phases given by the local phases of the laser beams. Suitable geometries in a square lattice yield a finite flux $\Phi$ through the plaquettes giving rise to the Harper-Hofstadter model.}
\label{Fig:renormalization}
\end{figure*}

\subsection*{Artificial gauge fields and topological band structures} 
When the driving scheme breaks time-reversal symmetry (TRS), it can generate a complex tunneling element $J=|J|e^{i\theta}$ with a Peierls phase $\theta$ \cite{Jimenez-Garcia2012, Struck2012} (Fig. 1b). TRS can be broken, e.g., by the multi-step scheme shown in Fig. 1b or by circular lattice shaking of a two-dimensional lattice \cite{Struck2013, Jotzu2014, Flaschner2016}. In a one-dimensional lattice, the Peierls phase leads to an effective band structure with a minimum shifted to a finite quasimomentum $\theta/d$, which again can be directly revealed via condensation of bosons at this minimum (Fig. 1b). When switching on the drive abruptly, one can also use the finite group veloctiy at quasimomentum zero of a BEC to obtain a net displacement with zero net force, know as quantum ratchet \cite{Salger2009}. Peierls phase are most relevant in two-dimensional lattices, where they give rise to artificial gauge fields and topological bands and therefore allow accessing a completely new class of important Hamiltonians for quantum simulation \cite{Dalibard2011, Lewenstein2012, Goldman2014, Zhang2018, Cooper2019}.

This can be understood as follows: In quantum mechanics, a particle with charge $e$ acquires an Aharonov-Bohm phase when encircling a magnetic flux. In the Peierls substitution on a lattice, this phase is attached to the tunneling elements and the sum of the Peierls phases around a plaquette yields the magnetic flux $\Phi$ through the plaquette according to $2\pi \Phi/\Phi_0=\sum_{\mathrm{Plaquette}} \theta_i$ with the magnetic flux quantum $\Phi_0=h/2e$. Using Floquet engineering, these Peierls phases are directly implemented independently of a real magnetic field. This means that the effects of a magnetic field on a charged particle become accessible for neutral particles such as cold atoms. Furthermore, one can easily imprint any value of the Peierls phases and thereby reach a magnetic flux quantum per plaquette, which would require a magnetic field of thousands of Tesla in a solid state crystal. For real magnetic fields, a central concept is the gauge freedom, which means that different vector potentials describe the same magnetic field. In Floquet engineering, however, one experimentally implements a specific gauge by directly imprinting certain Peierls phases on certain bonds around the plaquette. For this reason, one usually speaks of artificial gauge fields. This gauge choice of the Floquet protocol can in fact make a difference, e.g. in the adiabaticity time scale for ramping up a magnetic flux \cite{Wang2020}. The same considerations also hold for artifical gauge fields in bulk systems \cite{Dalibard2011}. 

Another way to induce Peierls phases is laser-assisted (or Raman-assisted) tunneling \cite{Jaksch2003, Gerbier2010}. In this scheme, the tunneling is suppressed by applying a tilt with an energy shift of $\Delta$ per lattice site and then resonantly restored by a Bragg-transition of two laser beams with wave vectors $\vec{k}_1, \vec{k}_2$ and frequencies $\omega_1, \omega_2$, where the resonance is ensured via the two-photon detuning $\delta \equiv \omega_2-\omega_1$ for $\hbar\delta=\Delta$ (Fig. 1c). The effect of the Bragg laser beams can also be viewed as a secondary moving lattice, which induces a time-periodic amplitude modulation at a lattice site $l$ at position $\vec{r}_l$ with locally varying phase $\phi_l=(\vec{k}_1-\vec{k}_2)\cdot \vec{r}_l$. The effective tunneling is described by a Bessel function of the first kind of order one $\mathcal{J}_1(K)$ and a Peierls phase on a bond between the sites $l$ and $l'$, which stems from the local phases of the two laser beams, i.e. the local phase of the amplitude modulation at the respective lattice sites as $\theta_{ll'}=(\phi_l+\phi_l')/2$. For a small driving amplitude $K$, where the Bessel function can be linearized, the tunneling element is therefore given by $K e^{i\theta_{ll'}}$ \cite{Eckardt2017}. 

In a two-dimensional square lattice, one can arrange the laser-assisted tunneling such that the Peierls phases yield a net magnetic flux $\Phi$, the value of which depends on the angle between the laser beams (Fig. 1c) \cite{Miyake2013, Aidelsburger2013, Kennedy2015, Aidelsburger2015, Tai2017}. This model is known as the Harper-Hofstadter model, a lattice version of the quantum Hall effect, which has topologically non-trivial bands with non-zero Chern number (Fig. 2a). Experiments have revealed the extended periodicity of the magnetic unit cell \cite{Kennedy2015}(Fig. 2b) and measured the Chern number of the lowest band via the transverse Hall drift in an accelerated lattice \cite{Aidelsburger2015}(Fig. 2c). Interestingly, this quantized response originally predicted for fermionic band insulators also appears for a homogeneous filling of the lowest band with thermal bosons. Finite Hofstader ribbons were also realized with one artificial dimension formed by internal spin states of the atoms, which obtain a coupling with spatially dependent Peierls phases by replacing the laser-assisted tunneling with Raman transitions within the same lattice site \cite{Celi2014}. The artificial dimension has sharp edges, which allows for the observation of the skipping orbits at the edges \cite{Mancini2015,Stuhl2015,An2017,Chalopin2020}. 

Another important model is the Haldane model on the honeycomb lattice \cite{Haldane1988}, which also has Peierls phases, but no net magnetic flux (Fig. 2d). It can be realized by elliptical lattice shaking \cite{Jotzu2014, Flaschner2016} or similarly in graphene sheets by illumination with circularly polarized Terahertz radiation \cite{Oka2009}, as well as in helically propagating photonic wave guides \cite{Rechtsman2013}. Experiments have mapped out the phase diagram via the closing of the band gap \cite{Jotzu2014}(Fig. 2e) as well as the Chern number via the quantized response in circular dichroism spectroscopy between the bands \cite{Tran2017,Asteria2019}(Fig. 2f).

Topological systems are an active area of current research and many models and phenomena are explored with cold atoms using Floquet engineering  \cite{Dalibard2011, Lewenstein2012, Goldman2014, Zhang2018, Cooper2019} such as non-Abelian gauge fields and synthetic spin-orbit coupling both in theory \cite{Ruseckas2005, Hauke2012, Burrello2013, Galitski2013} and experiment \cite{Wu2016, Li2017, Huang2018, Sugawa2018, Song2019}. Cold atom research has developed various new detection techniques complementary to those of solid-state physics, which allow directly revealing fundamental topological concepts such as Berry phase \cite{Duca2015}, Berry curvature \cite{Flaschner2016}, skipping orbits \cite{Mancini2015,Stuhl2015}, quantized Thouless pumping \cite{Nakajima2016, Lohse2016}, as well as completely new concepts such as linking numbers \cite{Flaschner2017, Wang2017, Sun2018, Tarnowski2019, McGinley2019} or Hopf invariants \cite{Unal2019}, which appear in quench dynamics. Current efforts aim at understanding the interplay of topology and interactions \cite{Regnault2011, Grushin2014, Vanhala2016, Stenzel2019, Rachel2018} and the challenges in the context of Floquet realizations of the latter \cite{Anisimovas2015, Qin2017, Plekhanov2017}.

\begin{figure*}%[t]  
\includegraphics[width=0.95\textwidth]{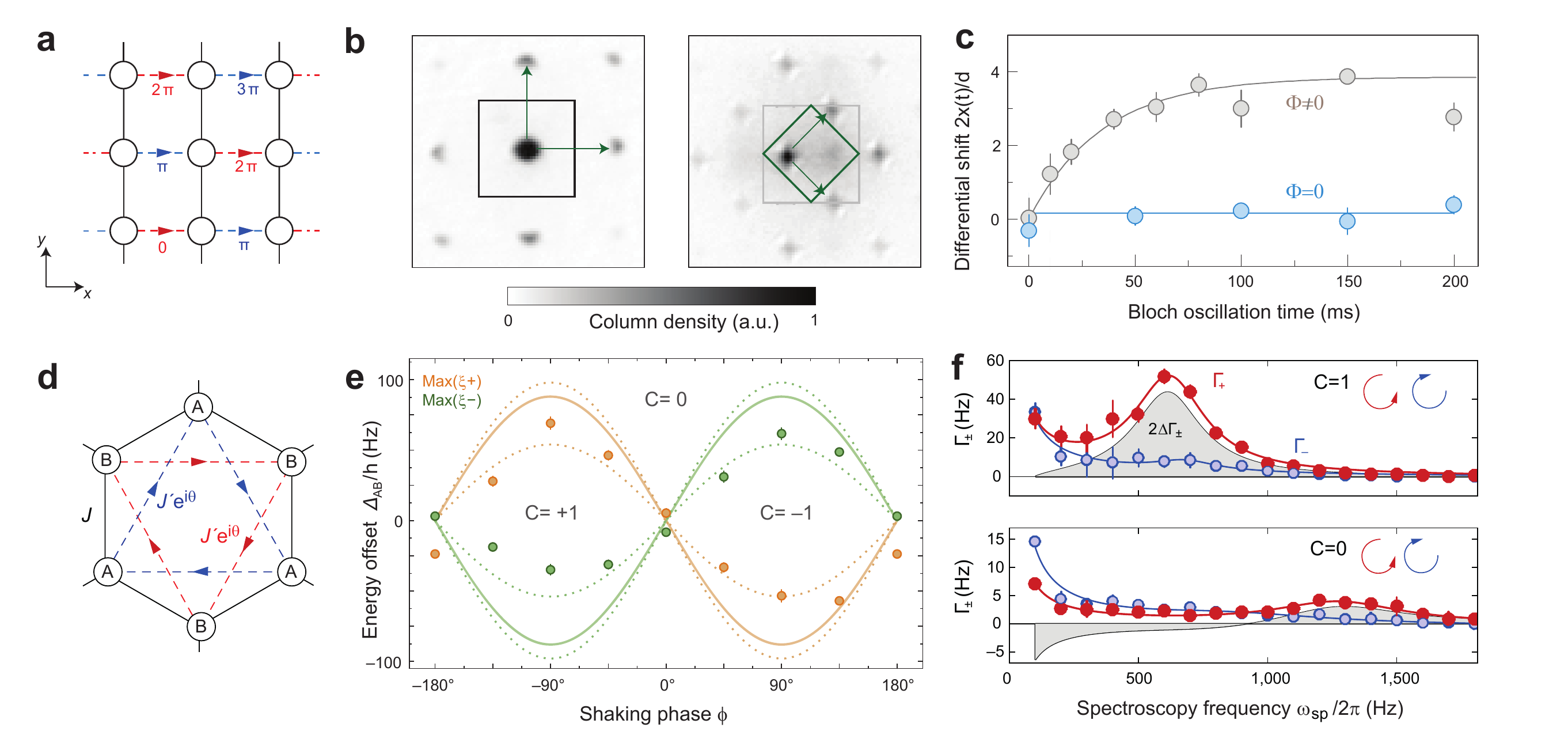}
\vspace{-0.cm}\caption{\textbf{Artificial gauge fields and topological band structures.} \textbf{a}, Harper-Hofstadter model. Square lattice with Peierls phases on the horizontal bonds imprinted via laser-assisted tunneling realizing a flux $\Phi=\pi$. The Peierls phases have a larger periodicity than the lattice and produce a larger magnetic unit cell. \textbf{b}, The quasi-momentum distribution of a BEC in the lattice without flux (left) and with flux (right) directly reveals this reduced symmetry. The squares denote the original and the magnetic Brillouin zone and the arrows denote the respective reciprocal lattice vectors. Adapted from \cite{Kennedy2015}. \textbf{c}, The effective bands have a non-trivial Chern number, which can be revealed by the transverse Hall drift in response to a lattice acceleration. The figure shows the differential center-of-mass shift $x(t)$ of the cloud as a function of the acceleration time along $y$ both with a magnetic flux (here $\Phi=\pi/2$, grey circles) and without magnetic flux ($\Phi=0$, blue circles). The drift is initially linear with the slope given by the Chern number. Adapted from \cite{Aidelsburger2015}. \textbf{d}, Haldane model. Honeycomb lattice with A and B sublattices and a staggered flux in the sub-plaquettes resulting from a Peierls phase $\theta$ on the next-nearest-neighbor tunneling element of strength $J'$.  \textbf{e}, The topological phase diagram with different Chern numbers $C$ as a function of the sublattice energy offset $\Delta_{\mathrm{AB}}$ and the phase $\varphi$ of the elliptical lattice shaking, which maps onto the Peierls phase $\theta$ in the Floquet realization. The topological phase transitions are indicated by the gap closing between the two lowest bands measured by Landau-Zener transitions. Adapted from \cite{Jotzu2014}. \textbf{f}, Chern number of the Haldane model for two different parameters measured via quantized circular dichroism, i.e. via the area under the difference of the depletion rate spectra $\Gamma_{\pm}(\omega_{\mathrm{sp}})$ obtained by additional circular shaking with $\omega_{\mathrm{sp}}$ with the two chiralities $\pm$. Adapted from \cite{Asteria2019}.}
\label{Fig:topology}
\end{figure*}

\subsection*{Floquet schemes in correlated systems}
A completely new class of Hamiltonians can be accessed by driving correlated Hubbard systems involving an on-site interaction energy $U$ \cite{Lewenstein2012}. In laser-assisted tunneling, the resonance condition for restoring tunneling in a tilted lattice now becomes dependent on the occupation of the lattice sites involved, i.e. dependent on whether the initial and final states involve the interaction energy $U$. For example starting from one atom at both sites, a tunnel process in the lattice tilted by $\Delta$ becomes resonant for a two-photon detuning of $\hbar \delta = \Delta+U$ (Fig. 3a). This allows restoring the tunneling processes for different occupations with separate pairs of laser beams and therefore to address them separately and imprint occupation-dependent (or density-dependent) Peierls phases. Such processes have been proposed as building block for many interesting exotic Hamiltonians. The density-dependent Peierls phases can, e.g., be mapped onto a one-dimensional anyon-Hubbard model with the Peierls phase becoming the statistical exchange phase of the effective anyons \cite{Keilmann2011, Greschner2015}(Fig. 3a). The same processes can be realized by restoring the tunneling in the tilted lattice via multicolor lattice shaking \cite{Strater2016} or amplitude modulation \cite{Ma2011, Cardarelli2016}. For an appropriate choice of the laser beams, the laser-assisted tunneling can additionally flip the spin of the atoms \cite{Bermudez2015}. This was employed in an experimental demonstration of density-dependent tunneling in a fermionic Mott insulator \cite{Xu2018}. Another possibility to obtain density-dependent tunneling is to periodically modulate the interaction strength itself taking advantage of Feshbach resonances. This strategy was employed in an experimental demonstration of correlated tunneling processes in a bosonic Mott insulator \cite{Meinert2016}. 

Density-dependent tunneling processes are essential for the implementation of dynamical gauge fields and lattice gauge theories \cite{Tagliacozzo2013, Wiese2013, Zohar2015}, which include a feedback of the neutral matter onto the synthetic gauge fields. Recent experiments have realized the first steps in this direction by implementing density-dependent tunneling processes in isolated double wells using sophisticated shaking schemes, e.g. involving two spin-states and spin-selective tilts \cite{Gorg2019, Schweizer2019}. Density-dependent gauge fields have also been realized by combining lattice shaking and modulation of the interactions \cite{Clark2018}. Such a combination of modulations allows engineering a broad class of unconventional Hubbard models with correlated tunneling \cite{Greschner2014,Zhao2019, Wang2020a}.

Periodic driving can also be employed to modify the spin interactions in Hubbard models (Fig. 3b). Using two spin states in a regime of a Mott insulator with one atom per site, spin interactions arise as a super-exchange process consisting of two virtual tunneling processes with tunneling element $J$, where the intermediate state with both atoms at the same lattice site is detuned by $U$. The superexchange coupling is therefore given by $J_{\mathrm{ex}}\sim J^2/U$ and renormalizing $J$ and $U$ will change $J_{\mathrm{ex}}$ accordingly. High-frequency lattice shaking will only reduce $J$ and therefore $J_{\mathrm{ex}}$. However, lattice shaking with $\Omega$ in near resonance with $U$ will renormalize the interaction to $U_{\mathrm{eff}}=U-\hbar\Omega$ and therefore reduce, enhance and even reverse the spin interactions depending on its frequency \cite{Coulthard2017}. A recent experiment has measured both the effect on the superexchange dynamics in isolated double wells \cite{Gorg2018}(Fig. 3b) as well as the resulting spin correlations in a fermionic many-body system on a honeycomb lattice changing from antiferromagnetic to ferromagnetic \cite{Gorg2018, Sun2019b}.  

\begin{figure*}%[t]
\includegraphics[width=0.6\textwidth]{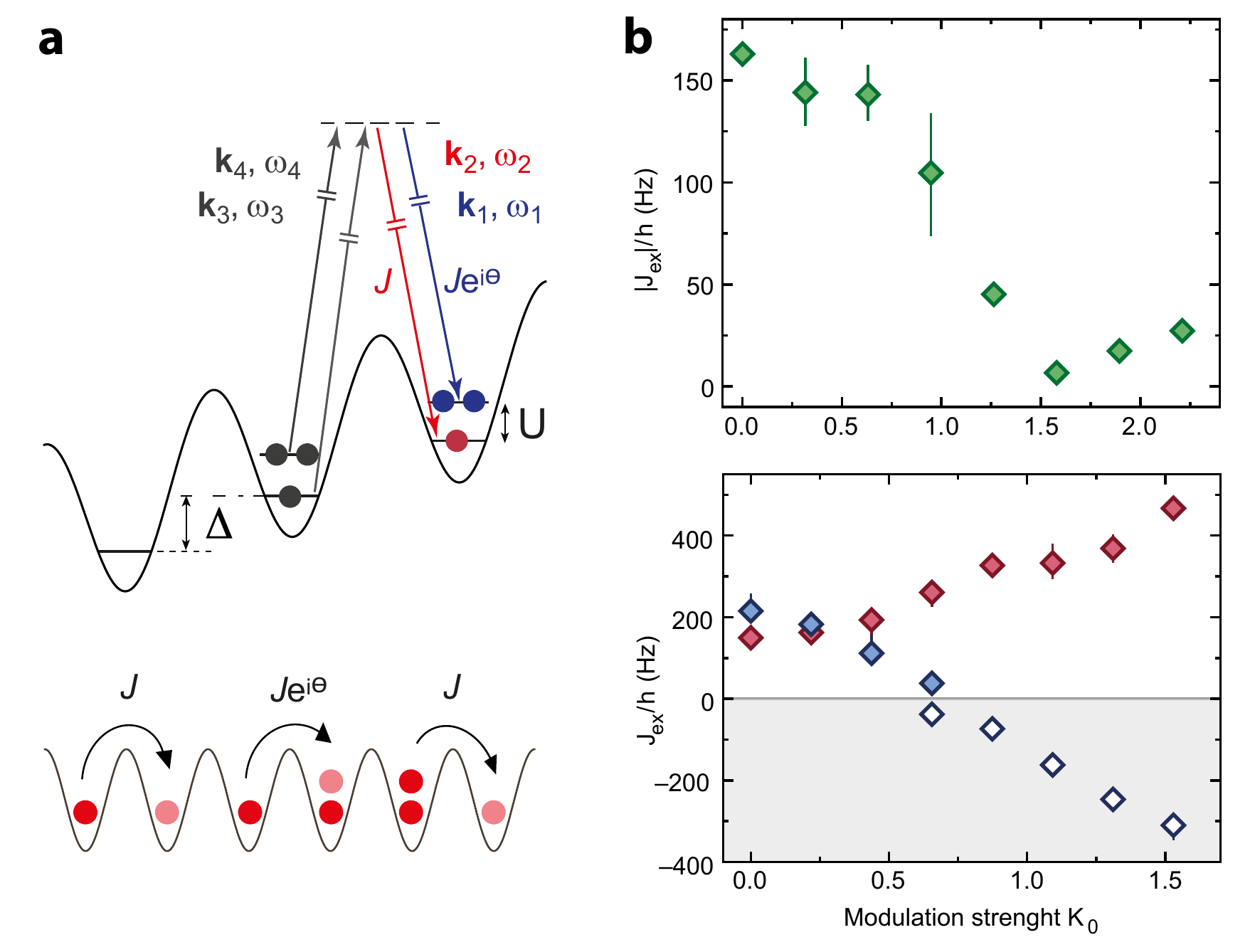}
\vspace{-0.cm}\caption{\textbf{Floquet schemes in correlated systems.} \textbf{a}, Top: Laser-assisted tunneling in presence of an on-site interaction $U$ allows selectively restoring tunneling processes depending on the site occupations giving rise to density-dependent Peierls phases. Bottom: Resulting tunneling processes, which carry a Peierls phase $\theta$ if the target site is already occupied. The two consecutive processes leading to an exchange of two particles therefore yield the phase $\theta$, which can be interpreted as an anyonic exchange phase, mapping the bosonic atoms to anyonic particles. \textbf{b}, Floquet engineering of the spin interactions in double wells of a fermionic Hubbard system. Top: Lattice shaking in the high frequency limit ($\hbar \Omega>> U$) leads to the expected renormalization of the superexchange coupling $J_{\mathrm{ex}}\sim J^2/U$ due to the renormalization of $J$. Bottom: For a driving in near resonance with $U$, the behavior is radically different: red-detuned driving ($\hbar\Omega < U$, red diamonds) enhances the superexchange coupling $J_{\mathrm{ex}}$ for increasing driving amplitude $K_0$, while blue-detuned driving ($\hbar\Omega>U$, blue and white diamonds) turns $J_{\mathrm{ex}}$ negative for larger driving amplitudes. Adapted from \cite{Gorg2018}.}
\label{Fig:correlated}
\end{figure*}

\subsection*{Physics beyond an equilibrium description} 
The physics of driven systems is richer than what is captured by the effective Hamiltonian. While the slow dynamics of Floquet systems can often be mapped to static Hamiltonians, Floquet systems are inherently non-equilibrium in nature with new properties beyond static concepts and they possess new phases without a static counterpart. These situations are of particular interest in nonequilibrium quantum statistical physics and they can be assessed with cold atoms. 

To get some insight into these issues, let us start with the Floquet theorem for time-periodic Hamiltonians, which states that the eigenstates are Floquet states $\ket{\Psi_n(t)}$ in analogy to Bloch states for spatially periodic potentials: they can be written as a time-periodic part times a phase evolution $\ket{\Psi_n(t)}=\ket{u_n(t)}e^{-i\epsilon_n t/\hbar}$ with the Floquet modes $\ket{u_n (t+T)} =\ket{u_n (t)}$. The states do not have energies, but quasi-energies $\epsilon_n$, which are defined modulo $\hbar\Omega$, because the system can always exchange energy quanta of $\hbar\Omega$ with the drive: Energy is not conserved in a driven system. In the extended zone scheme, this leads to many copies of the spectrum and correspondingly to new band gaps (see Fig. 4b). These new band gaps in the quasi energies can have important consequences such as new collision processes becoming resonant \cite{Bilitewski2015, Reitter2017}.

An illustrative example of the consequences of the Floquet band gaps is the anomalous Floquet topological insulator, where the Floquet-nature of the phase gives rise to new topological properties by redefining the connection between the bulk topological index (e.g. the Chern number) and the existence of chiral edge states. According to the bulk-edge correspondence of static systems, chiral edge states appear for non-trivial bulk bands. But in the Floquet system, one can have anomalous edge states, which appear in a system with zero Chern number \cite{Kitagawa2010, Rudner2013, Quelle2017}. The game changer comes from the additional gap between bands around the Brillouin-zone of quasi-energies (see Fig. 4a,b). The formal description requires the introduction of winding numbers to characterize the gaps and not surprisingly, the calculation of these winding numbers cannot be done from an effective static Floquet Hamiltonian, but requires to take into account the full time-dependency of the system \cite{Rudner2013}. Theses states have been realized with photonic waveguides \cite{Mukherjee2017, Maczewsky2017} and recently with cold atoms \cite{Wintersperger2020}, which promise access to bulk and edge observables in the same system. This example shows that established principles have to be reexamined in a non-equilibrium context. In fact, a new classification of topological insulators has been introduced for Floquet systems \cite{Potter2016, Roy2017}.

A consequence of the absence of energy conservation in driven systems is that in the long-time limit the system will heat up to unconstraint temperature \cite{DAlessio2014}. However, there is a prethermal regime at an intermediate time scale, where the system is in an equilibrium-like state and which can be employed to study the physics of the effective Floquet Hamiltonian \cite{Bukov2015b, Mori2018, Seetharam2018}. The time scale of this prethermal regime, which can be up to $10^4$ driving periods, depends critically on drive frequency and interaction strength and generally increases for high-frequency driving. Therefore experimental studies with cold atoms have investigated heating in various conditions \cite{Boulier2019, Wintersperger2020b} and mapped out the parameter space to identify and characterize the prethermal regime, in particular for bosons in driven optical lattices finding exponentially suppressed heating rates \cite{Messer2018, Singh2019, Rubio-Abadal2020}(Fig. 4c). 

\begin{figure*}%[t]
\includegraphics[width=0.5\textwidth]{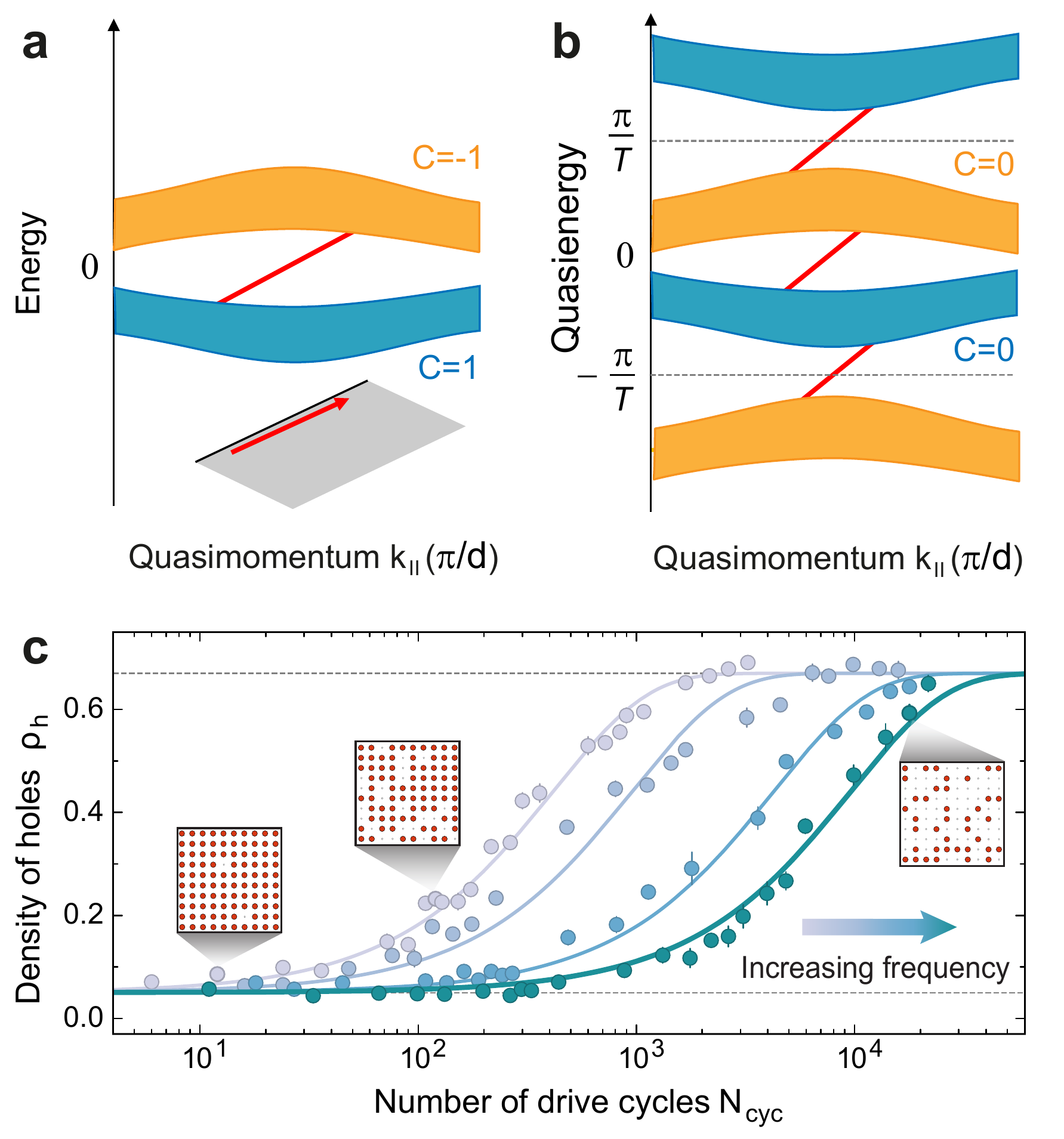}
\vspace{-0.cm}\caption{\textbf{Physics beyond an equilibrium description.} \textbf{a},\textbf{b} Illustration of the anomalous Floquet topological insulator. Energy spectrum of a two-dimensional lattice in a strip geometry as a function of the quasimomentum $k_{\parallel}$ parallel to the strip (\textbf{a}). In the case of the driven system (\textbf{b}), the energy is replaced by a quasienergy and the spectrum contains many copies of the bands as well as new gaps at $\pm \pi/T$. The bulk-edge correspondence makes a connection between the topological index of the bulk bands (here the Chern number $C$) and the existence of chiral edge states at the edges of the system (red arrow in the inset), which lie in the bulk band gaps (red lines): the Chern number is given by the difference of the outgoing and ingoing edge states. In a static system (top), edge states can only occur, when the bulk bands have non-zero Chern number. In a Floquet-system with its additional gaps, one can have two chiral edge states despite the zero Chern numbers of the bulk bands. These edge states are therefore called anomalous. \textbf{c}, Floquet thermalization in a driven interacting Bose-Hubbard system. Measurement of the density of holes as a function of the driving cycles $N_{\mathrm{cyc}}$ measured after ramping the system adiabatically into the atomic limit of a Mott insulator. Heating processes revealed by the formation of holes set in after $10^2-10^4 N_{\mathrm{cyc}}$. Strikingly, the heating rate decreases for increasing shaking frequency $\Omega$, which is an indication of the exponential slow down of heating characteristic of Floquet prethermalization. Adapted from \cite{Rubio-Abadal2020}.}
\label{Fig:quasienergies}
\end{figure*}

\subsection*{Outlook} 
In this review article, we have summarized how Floquet engineering of quantum gases has evolved into a very active field of research exploring ever more complex systems. One ubiquitous challenge is the heating in the driven system, which has to be controlled. Therefore the recent systematic studies of the prethermal regime are an important benchmark. Current efforts aim at exploring ways to circumvent heating, be it by using non-ergodic systems \cite{Abanin2019, Potirniche2017} or by employing destructive interference of excitation paths via a two-color drive \cite{Viebahn2020}. Controlling heating would allow accessing strongly correlated phases such as fractional quantum Hall states \cite{Rachel2018} or half-integer Mott insulators \cite{Keilmann2011}, as well as extending quantum simulation to high-energy physical concepts such as full-fledged dynamical gauge fields \cite{Tagliacozzo2013, Wiese2013, Zohar2015}. Furthermore, Floquet systems allow accessing totally new systems such as discrete time crystals, which spontaneously break the discrete time symmetry of the Floquet system to a reduced symmetry with to a period, which is a multiple of the driving period \cite{Sacha2018}.

Floquet techniques have been introduced to ultracold atoms as a way to tailor the systems and to add new properties such as artificial gauge fields. After more than one decade of intense research and cross-fertilization between theory and experiments, we can say that Floquet techniques do much more than that. They have opened a new avenue to exotic states of matter and to exciting non-equilibrium physics and they will continue to play an important role.

\subsection*{Acknowledgements}
This work was supported by the European Research Council (ERC) under the European Union's Horizon 2020 research and innovation programme under grant agreement No. 802701 and by the Deutsche Forschungsgemeinschaft (DFG, German Research Foundation) via Research Unit FOR 2414 'Artificial gauge fields and interacting topological phases in ultracold atoms' under project number 277974659 and via the SFB 925 'Light induced dynamics and control of correlated quantum systems' under project number 170620586. 

%\bibliography{mybib}

%merlin.mbs apsrev4-1.bst 2010-07-25 4.21a (PWD, AO, DPC) hacked
%Control: key (0)
%Control: author (8) initials jnrlst
%Control: editor formatted (1) identically to author
%Control: production of article title (-1) disabled
%Control: page (0) single
%Control: year (1) truncated
%Control: production of eprint (0) enabled
%

\end{document}